\input harvmac.tex
\noblackbox
\Title{\vbox{\hbox{HUTP-96/A004}\hbox{\tt hep-th/9602022}}}
{Evidence for F-Theory}
\bigskip\vskip2ex
\centerline{Cumrun Vafa}
\vskip2ex
\centerline{\it  Lyman Laboratory of Physics, Harvard
University}
\centerline{\it Cambridge, MA 02138, USA}
\vskip .3in
We construct compact examples of D-manifolds for type IIB strings.
The construction has a natural interpretation
in terms of compactification of a
12 dimensional `F-theory'.  We provide evidence for a
more natural reformulation of type IIB theory in terms of F-theory.
Compactification of M-theory on a manifold $K$ which admits elliptic
fibration is equivalent to compactification of F-theory on
$K\times S^1$.  A large class of $N=1$ theories in 6 dimensions
are obtained by compactification of F-theory on Calabi-Yau
threefolds.  A class of phenomenologically
promising compactifications of F-theory is on
$Spin(7)$ holonomy manifolds down to 4 dimensions.  This may
provide a concrete realization of Witten's proposal for solving
the cosmological constant problem in four dimensions.
\Date{\it {Feb. 1996}}
\newsec{Introduction}
Discovery of string dualities has opened up a new era
in our understanding of string theory.  A global
picture of dualities has not yet emerged and it seems that
different dualities are most easily `understood'
from different views.  One successful viewpoint has been the notion
of M-theory \ref\sch{J. Schwarz,
hep-th/9508143, hep-th/9510086}\ref\witva{P. Horava and E.
Witten, hep-th/9510209}\ref\muda{K. Dasgupta and
S. Mukhi, hep-th/9512196}\ref\witan{E. Witten, hep-th/9512219}\ref\dufw{M.J.
Duff, R. Minasian and E. Witten, hep-th/9601036}.
Even though this has led to many insights into string dualities,
there are some cases that appear
less naturally in this formulation.
  For example to understand
the $SL(2,{\bf Z})$ invariance of type IIB in ten dimensions
we first have to compactify down to nine dimensions and compare with
$T^2$ compactifications of M-theory \sch\ref\aspi{P.S. Aspinwall,
hep-th/9508154.}\ where $SL(2,{\bf Z})$ gets interpreted as the
symmetry of the torus.  However,
 only in the limit where $T^2$ is of zero area do we obtain
type IIB in ten dimensions.  It would have been far more
satisfactory if the $SL(2,{\bf Z})$ invariance of type IIB
in ten dimension had a geometric meaning already in ten dimensions.

In this paper we face this question even more strongly
in consideration of compactifications of type IIB
string on D-manifolds \ref\bsvo{M. Bershadsky,
V. Sadov and C. Vafa, hep-th/9510225}.  So far it has
been difficult to obtain compact examples of such manifolds--
here we will find that if the p-branes are D-branes of different
$(p,q)$ strings \sch , one can construct compact examples.  In arguing for
the existence of these vacua a 12 dimensional viewpoint emerges.
One can have two views about this 12 dimensional origin
of type IIB:  either it is
an auxiliary manifold just useful for constructing vacua
of string theory involving D-manifolds
or it is more real.  In support of the latter interpretation,
which we call the `F-Theory',
we point out that this 12 dimensional viewpoint also solves another
puzzle of type IIB strong/weak duality:
The D-string, having a gauge field on its worldsheet has
a critical dimension of 12 and
actually lives
in $(10,2)$ space and is related to ten dimensional type IIB
theory by a null reduction \ref\ov{H. Ooguri
and C. Vafa, Mod. Phys. Lett. {\bf A5} (1990) 1389;
 Nucl. Phys. {\bf B361} (1991) 469;
Nucl. Phys. {\bf B367} (1991) 83.}.  The {\it
on shell} physical
states only carry ten dimensional momenta.  However off-shell
states may have more non-trivial dependence on the
extra two coordinates and may make them more real.

As concrete examples we consider compactification of F-theory
on $K3$ which can also be interpreted as type IIB compactification on a
D-manifold, $S^2$,
where we have turned on 24 7-branes on $S^2$, and argue that it is dual to
heterotic
string compactification on $T^2$.  We also consider compactifications
of F-theory to 6,5 and 4 dimensions on Calabi-Yau, $G_2$ holonomy
manifolds and $Spin(7)$ holonomy manifolds. Upon compactification
on a circle these would be related to M-theory compactifications
on the same manifolds.  The 6-dimensional
compactification of F-theory lead to new type II vacua
with $N=1$ supersymmetry with a number of tensor multiplets and gauge
multiplets. Compactifications of F-theory on
$Spin(7)$ manifolds down to 4 dimensions apparently
has no supersymmetry.  However upon compactification
to three dimensions it is related to a supersymmetric M-theory.
This may lead to a solution of
the cosmological constant problem along the lines proposed by Witten
\ref\witcos{E. Witten, Int. J. Modern Phys. {\bf A10} (1995) 1247;
hep-th/9506101}.

\newsec{New Vacua for Type IIB Strings}
The bosonic content of the type IIB consists of
$g_{\mu \nu},B^A_{\mu \nu},\phi$ from the NS-NS sector
and $\tilde \phi,B^P_{\mu \nu},A^+_{\mu\nu\alpha\beta}$
from the R-R sector.  Under the conjectured ten dimensional
$SL(2,{\bf Z})$ strong/weak duality \ref\hut{C.M. Hull and P.K. Townsend, Nucl.
Phys. {\bf B438} (1995) 109}\
the complex field $\tau=\tilde \phi + i {\rm exp}(-\phi)$
transforms in the same way as the modulus of a torus.  The
antisymmetric tensor fields $B^A$ and $B^P$ get exchanged
and the metric $g$ and the four form potential $A^+$ are invariant.

Let us consider vacuum solutions of type IIB for which
$B^A=B^P=A^+=0$.  We first consider solutions which leave
an 8-dimensional Lorentz group invariant.  Let $z$ denote
as a complex coordinate the 9-10 direction.
The relevant low energy lagrangian is
\eqn\act{S=\int \sqrt{g}\big(R
+ {\partial \tau \partial {\bar \tau}\over \tau_2^2}\big )}
Then we are
looking for solutions of low energy lagrangian where
$\tau (z,\bar z ) $ and $g_{z\bar z}(z,\bar z)$
vary over $z$ and the
rest of the components of $g$ are flat.  Moreover we look
for solutions where half of the supersymmetries are preserved.
This BPS condition implies that $\tau$ is only a
function of $z$ (or $\bar z$).  An example of this is the vacuum for
 a Dirichlet $7$-brane, where the worldvolume of the 7-brane
coincides with the uncompactified 8 dimensional spacetime.
Let the 9-10 position of the $7$-brane be at $z=0$.  Since
a 7-brane is a magnetic charge for the $\tilde \phi$ field,
it implies that as we circle $z=0$, we have $\tau \rightarrow \tau +1$.
Together with the holomorphy of $\tau$ we learn that
near $z =0$
$$\tau(z) \sim {1\over 2\pi i}{\rm log}(z)$$
Near $z=0$ the above adiabatic description of the action
breaks down. However that is precisely the regime where
we can use perturbative string theory and we know that
the above solution is consistent with having a Dirichlet
7-brane at the origin and so is an acceptable solution.
However the above ansatz runs into infrared difficulty for large $z$.
Viewed as a `cosmic string' the energy per unit length of this solution
diverges which means that the equation for gravity cannot
be solved in a consistent way. This is exactly the same
problem encountered and solved in the context of `stringy cosmic strings'
\ref\scs{B. R. Greene, A. Shapere, C. Vafa
and S.-T. Yau, Nucl. Phys. {\bf B337} (1990) 1}.
In fact the connection between 7-branes of type IIB and stringy
cosmic strings has already been pointed out
in \ref\gp{G. Gibbons, M.B. Green and M.J. Perry, hep-th/9511080}.  In the case
of stringy cosmic string
one was considering  a toroidal compactification  which
led to a complex moduli $\tau$.  To construct finite energy
solutions where $\tau$ depends holomoprhically on
a complex parameter $z$ it was found that it is crucial to use the ambiguity
of $\tau$ to allow $SL(2,{\bf Z})$ jumps in $\tau$.

Once we allow jumps up to an $SL(2,{\bf Z})$ we can
construct a solution. To get a compact space we need to wrap
the complex plane $24$ times around the fundamental domain
of $SL(2,{\bf Z})$. In the case considered in \scs\ this
gives rise to the four dimensional $K3$ compactification
when we take into account the `hidden' torus.  In fact
the existence of the hidden torus was crucial in \scs\
to argue for the existence of this solution.  In particular
in regions were $\tau_2\rightarrow \infty$ the fact that
the total space, including the $T^2$, is not singular,
was crucial in establishing the consistency of the solution.
Here we can use the same solution as there, except that now
we seem to have no `hidden' torus!
The analogy is so strong that we are led to search for a hidden
torus and in fact in the next section we present evidence for its
existence.  However, in this section let us continue our discussion
of the properties of the 8-dimensional solution we have just constructed.

To make the solution more concrete let us specify how
$\tau$ depends on $z$.  Consider the torus given by
$$y^2=x^3+f^8(z)x+f^{12}(z)$$
where $f^n$ are polynomials of degree $n$ in $z$.
The above equation defines a torus as a function of $z$.
The
ratio of the degrees of the two polynomials  is set by the
condition that as $z\rightarrow \infty$ we get a non-singular
torus.
Moreover the total degree of each is set by the
condition that we wrap the $z$-sphere 24 times around
the complex moduli of $T^2$.  To see this note that
the points where the torus degenerates corresponds
to where the discriminant of the cubic vanishes, i.e. when
$$27 (f^8)^3-4 (f^{12})^2=0$$
which is a polynomial of degree 24, and thus has 24 solutions.
Note that the number of independent variables in the above
solution is 18: 9 coefficients from $f^8$ plus 13 coefficients from
$f^{12}$, minus 3 from $SL(2,C)$ action on the $z$-plane minus
1 from the fact that scaling $f^8\rightarrow \lambda^2 f^8$ and
$f^{12}\rightarrow \lambda^3 f^{12}$ does not change the torus.
On the other hand, the solution in \scs\ allows for two further
moduli:  One for the size of the $S^2$ and the other for the
size of $T^2$.  For us, the size for $T^2$ is not dynamical
so that is not a moduli.  However the size of $S^2$ is
observable.  Note that we {\it cannot} turn on $B^A$ or $B^P$.
Not even to set them to a constant, because we are  using $SL(2,{\bf Z})$
transformations and they are not invariant under it.  So they are
frozen to be zero.  Thus in particular the Kahler moduli of $S^2$
is a real parameter and is not complexified as is customary in more
conventional string backgrounds.
 So the total moduli is $18$ complex parameters
describing the complex moduli of the elliptic fibration over $P^1$
and one real parameter describing the size of $S^2$.
Note that we have managed to find a {\it compact} manifold
where we have a condensation of D-branes. This is the
first compact example of D-manifolds proposed in \bsvo.
Note that unlike open strings where condensation of D-branes is
very natural \ref\pod{J. Dai, R. G. Leigh and J. Polchinski, Mod. Phys. lett,
{\bf A4}
(1989) 2073}\ref\hor{P. Horava, Phys. Lett. {\bf B231} (1989) 251}, for closed
strings
the story was not so clear. In fact there was a kind of `no-go theorem'
which runs as follows:  If we have a D-brane in a compact manifold
then the flux of the D-brane charge has nowhere to go, and so
we would run into inconsistencies.  What has happened in
the example constructed above is that we have used the non-abelian
nature of the relative D-brane charges coming from different $(p,q)$ strings
related to each other by $SL(2,{\bf Z})$ to avoid any inconsistencies.

Let us see how many gauge fields we will get in this compactification.
The fact that we have twenty four 7-branes, and that each one comes with
a $U(1)$ gauge field, might at first suggest that we have generically
a $U(1)^{24}$.  This however cannot be the case.  The easiest
way to see this is that the condensation of
scalars which are  in the adjoint of
the $U(1)$'s are responsible for changing
the relative position of the 7-branes. However above we found
that there are at most 18 relative positions (in the 9-10 plane)
free to be changed.  Thus we have $U(1)^{18}$ plus another $U(1)^2$
coming from the overall complex shift of all the 7-branes
 which are the graviphotons.

This reduction of the number of $U(1)$ gauge fields from a naive
counting of the D-branes deserves further comments.  Even though we
have twenty four 7-branes, they are not perturbative D-branes of a given
string theory.
More generally they are the perturbative D-branes of some $(p,q)$
string, because we have used $SL(2,{\bf Z})$ in constructing
our solution.  Near a region where we have a 7-brane of a $(p,q)$
theory we can use perturbative string description of the $(p,q)$ theory
to study the structure of string spectrum. However in going from
one $(p,q)$ string spectrum to another, we may be double
counting a given state.  In the case at hand it turns out that we
can get at most 18 of the D-branes to be the 7-branes of a given theory.
This is consistent with the fact that the other $U(1)$'s are linear
combinations of these 18 $U(1)$'s.

Note that in addition to the gauge fields, we have $A^+$ 4-form.
We can either take two component to be the volume element on $S^2$,
leaving us with a 2-form in 8 dimensions,
or we can take it to be the uncompactified components which is dual
to a 2-form in 8 dimensions.  Together with the fact that
$A^+$ has a self-dual field strength, we thus end up with
one antisymmetric two form in eight dimensions.  Putting all the
spectra and the number of supersymmetries together we see that
we have exactly the same spectrum as the toroidal compactification of
the heterotic string from 10 to 8 dimensions.  We conjecture
that our solution is dual to it, where the role of the coupling
of heterotic string is played by the size of $S^2$.  Weak coupling
heterotic string corresponds to small $S^2$ and large coupling corresponds
to big $S^2$.

One simple check for this identification is the fact that the
field strength of the antisymmetric field in 8 dimensions, $H$,
in both cases satisfies
$$dH={1\over 2}[ p_1(R)-p_1(F)]$$
where $p_1(R)$ (resp. $p_1(F)$) refer to first pontryagin
class of $R$ (resp. $F$).
In the case at hand
such a correction is present.  To see this note that
$H$ is the field strength dual to the RR gauge potential
$A^+$ in eight dimensions and view the 8 dimensional
spacetime as the worldbrane of the 7-branes.  Then
as argued in \ref\bsvt{M. Bershadsky, V. Sadov and C. Vafa,
hep-th/9511222}\ in the worldbrane of
the D-branes there are such corrections (and we need exactly
24 7-branes to get the coefficient of $p_1(R)$ to come out correctly).
That there are such corrections was deduced implicitly in \bsvt .
At least for the $p_1(F)$ term a direct computation of this
has been performed \ref\li{M. Li, hep-th/9510161}\ref\calee{C.G. Callan, C.
Lovelace, C.R. Nappi and S.A. Yost,
Nucl. Phys. {\bf B308} (1988) 221}\ref\doug{M.
Douglas, hep-th/9512077}.

Let us give two additional arguments in favor of this conjectured
duality\foot{
The first one was pointed out to us by N. Seiberg.}.  Consider
the compactification of M-theory on $K3$ which
are themselves $T^2$ fibration over $S^2$.  Since
compactification of M-theory on $T^2$ is equivalent to
compactification of type IIB theory on $S^1$, to the extent
that we can trust the adiabatic argument \ref\vw{C. Vafa and
E. Witten, hep-th/9507050},  we
end up on type IIB side with a compactification on $S^1\times S^2$
where $\tau$ varies over $S^2$ as in the above.  So we see that
the compactification of the above solution on a further $S^1$
is on the same moduli as M-theory on $K3$ which is also
conjectured \ref\wits{E. Witten, hep-th/9503124
}\ to be on the same
moduli as $T^3$ compactification of heterotic string.

The second argument is as follows: If we compactify further
on $T^2$ and use $T$-duality once on each of the two circles
of $T^2$ we end up again with type IIB, but now instead of
twenty four 7-branes we have twenty four 5-branes.  In
fact it was shown locally \ref\ogv{H. Ooguri and
C. Vafa, hep-th/9511164}\bsvo\
that type IIB compactified on a D-manifold with 24 5-banes
is dual to type IIA compactified on $K3$ which is itself
conjectured to be dual to heterotic string on $T^4$ \hut \ref\cv{C. Vafa,
unpublished.}.  Here we have managed to find a concrete global realization
of this proposal\foot{
Note also that we can also deduce that if we compactify
only on one $S^1$ and use T-duality on that circle we have
a consistent background of type IIA with 24 6-branes, which is equivalent
to M theory on K3.}.  Recall that in \bsvo\ the identification
of the behaviour of type IIA compactification
near an $A_{n-1}$ singularity of K3 with type IIB on $n$
symmetric 5-branes \ogv\ coupled with strong/weak duality
of type IIB relating this to $n$ coincident Dirichlet 5-branes
predicted the appearance of an enhanced $SU(n)$ gauge symmetry.
It is natural to ask what happens to other enhanced gauge symmetries
such as $E_8$ and how it is consistent with the global
picture of the D-manifold we have developed above.

Let us consider this in a little more detail.  The local singularity
of $E_8$ is given by
$$x^5+y^3+z^2=\mu$$
If we deform by a polynomial in $y$ we get two $A_4$
singularities each of which corresponds to 5 7-branes
of one $(p,q)$ string coming together.  So we can predict
 two $SU(5)$ gauge symmetries.
  If we deform by a polynomial in $x$
we get four $A_2$ singularities each of which corresponds to
3 7-branes of some $(p',q')$ string coming together.  So again
we can predict four $SU(3)$ gauge symmetries.
If we do not deform at all we cannot predict what gauge
symmetry we have because we have a region where there
is no $(p,q)$ string  in which the coupling is weak.
In this way we avoid the contradiction of having
to see an $E_8$ gauge symmetry in terms of $D$-branes.
This is somewhat analogous to the avoidance of similar
contradictions for type I-heterotic duality \ref\powi{J. Polchinski and
E. Witten, hep-th/9510169}.

Here we have proposed that heterotic theory
compactified on $T^2$ has a strong coupling
regime which behaves like a 10 dimensional type IIB string
with some 7-branes. Note however that if we wish
to describe it in terms of a weak coupling limit of some
string theory we will at most get an $A_n$ type gauge symmetry.
  On the other hand type I-heterotic
duality similarly suggests a decompactified type I string \wits ,
which are somewhat analogous to the case at hand except that
in such cases the weak coupling regime will give only a $D_n$
type gauge symmetry.

\newsec{Type IIB and F-theory}
In the above 8 dimensional example we have seen how crucial
the modulus $\tau$ has been in constructing a solution and
in some sense it is convenient to think of it as describing
the complex moduli of a  $T^2$, in which case we can
think of the above 8 dimensional solution as `compactification'
from 12 dimensions to 8 dimensions, on a $K3$.  Of course
this $K3$ has to admit an elliptic fibration with a frozen
Kahler class for the torus.
With the lesson of string dualities in geometrizing various
symmetries the idea of a `real' $T^2$ is very appealing.
  In fact this can be
partially realized as follows \sch \aspi :  Consider M-Theory
on $T^2$.  This theory has an $SL(2,{\bf Z})$ symmetry as a geometric
symmetry of the torus.  It also has a scale corresponding to the
area of $T^2$.  If we take this are to zero, and use the relation
between M-theory and type IIA, and use T-duality to relate
to type IIB one can see that we will end up with type IIB theory
in 10 dimensions in this limit.  Even though this partially
geometrizes the $SL(2,{\bf Z})$ of type IIB, the
zero area limit we are taking
is singular and in this limit the geometric description of
M-theory breaks down.  One is left with a desire
to explain this $SL(2,{\bf Z})$ in a more direct fashion
on the type IIB side.  One would think that there
should be a dual to type IIB, just as M-theory is dual to type
IIA.

In fact we will see now that
 one can use the strong/weak duality above to {\it
derive} the existence of a 12 dimensional theory with signature (10,2).
The strong/weak duality of type IIB,
together with the observation of Polchinski \ref\jpo{J. Polchinski,
hep-th/9510017}
that D-branes
carry RR charge, implies that the dual of fundamental type IIB
string which couples to $B^A_{\mu\nu}$ is a Dirichlet 1-brane, the D-string,
which couples to $B^P_{\mu\nu}$.  The modes on the Dirichlet
1-brane worldsheet are the same as one expects for type II string
theory, namely the dimensional reduction of 10d YM to 2d.
However, there is an extra mode: the $U(1)$ gauge field which lives on the
worldbrane seems apriori to have no analog
for the worldsheet of the type IIB string
theory.  Thus we are faced with quantizing a string
with $N=1$ worldsheet supergravity together with $N=1$ super-Maxwell.
  This exact problem was in
fact already considered a while back \ov\ in
constructing heterotic versions of
 type II strings and $N=2$ strings, the relevant features
of which we will now review. We will not need, however, the
coupling of type II string to $N=2$ strings and shall mainly
concentrate on a reformulation of type IIB strings.

If we introduce on the worldsheet of type IIB strings
a $U(1)$ super-Maxwell field the critical dimension
of the theory changes, because now we have to introduce
additional ghosts of spin (0,1) which shifts the central
charge by $-2$.  So the critical dimension is now $10+2=12$.
Moreover the addition of new ghosts implies that the signature
of the additional space is (1,1), giving the total space a signature
of (10,2).  Consider $v^\mu$ a vector in the $(10,2)$ space.
Then the $U(1)$   current couples to $v_\mu DX^\mu$.
For BRST invariance it turns out that $v\cdot v=0$, i.e.,
$v$ is a null vector. Choose coordinates so that $v$ lies
in a $(1,1)$ subspace $M^{1,1}$.
  The BRST
invariance of the states implies a null
reduction of the theory and thus the physical states
carry no momentum or oscillation along $M^{1,1}$.  Thus the
physical states are the same as that of the type IIB theory.
So this is consistent with D-string being dual to type IIB
strings which until now lived in 10 dimensions.  However
we would like to take the existence of these two extra dimensions
a bit more seriously as some kind of `off-shell' states.
For this purpose it is convenient to compactify the $M^{1,1}$
part of the space. If we compactify left- and right-movers
independently we get a Narain lattice
$$\Gamma^{1,1}_L\oplus \Gamma^{1,1}_R$$
where the subscripts $L,R$ refer to left- and right-moving
momenta.  Of course this can more generally be deformed so
that the lattice does not have a product structure between left and
right.  The most general lattice we will get in this way is a lattice
$$\Gamma^{1,1}_L\oplus \Gamma^{1,1}_R\rightarrow \Gamma^{2,2}$$
which is the {\it same} Narain lattice as one would get upon
compactification on a {\it Euclidean} $T^2$.  The main difference
is that we are exchanging one left-moving string mode with one
right-moving one in interpreting the momenta for the compactification
of $(10,2)$ theory on a $(1,1)$ space.
At any rate the moduli of a $\Gamma^{2,2}$ lattice,
however we may wish to interpret the individual components,
is parametrized
by two complex parameters, which can be identified as the
complex structure $\tau$ and the Kahler structure $\rho$ of
a Euclidean torus.  Even though the BRST invariance
rules out non-zero momentum states in the internal torus,
the zero modes are not ruled out.  We conjecture that the zero mode
corresponding to one of the two moduli, say $\tau$ is physical
and is to be identified with the ($\phi, \tilde \phi$) of type IIB.
The fact that only one of the two moduli is physical is not
so strange and is the case for $N=2$ strings where only one
of the two moduli is dynamical. For example $N=2$ string amplitudes
on $T^2\times R^2$ depends only on $\tau$ and not on $\rho$
\ref\ova{H. Ooguri and C. Vafa, hep-th/9505183}.  This suggests that even
though
we are dealing with an internal theory with signature $(1,1)$
we can view it geometrically as if we are compactifying it
on a {\it Euclidean} $T^2$ whose only dynamical degree of freedom is
its complex structure.  This is the version of F-theory we will
be mostly using in this paper.

To make the above idea more concrete one will need to recall
certain facts about $N=2$ strings.
In the context of $N=2$ strings, it was proposed
\ov\ that there should be a more symmetrical reformulation of the theory
where the fundamental degree of freedom instead of
strings with a (1,1) worldsheet is an object with 4 dimensional
worldvolume with signature (2,2).  Since the same reasoning
applies here let us briefly review the idea of \ov .  $N=2$ strings
has a 4 dimensional target, but surprisingly has only one particle
in its spectrum.  In particular the infinitely many oscillatory
modes of strings are not physical states. This is a bit strange
because worldsheet is 2 dimensional and the target is 4. What happens
to the normal oscillations?  Technically they are gotten rid
of by the contributions of $U(1)$ current in the BRST invariance
of physical states.  This suggests that there may be another formulation
where the worldvolume is 4 dimensional in which case the absence of oscillatory
modes will be more manifest.  In fact if we just view string as a
`geometrization' of Virasoro algebra, we will also be led to the
same conclusion.  From the algebraic point of view
 the coordinates of a
Riemann surface arises by considering evolution operator
$$z^{L_0}$$
Similarly in the $N=1$ super-Virasoro we add an odd variable $\theta$
and consider the evolution operator
$$z^{L_0}{\rm exp}(\theta G_0)$$
the $(z,\theta)$ is then geometrically interpreted as the coordinates
of a super-Riemann surface.  Now, if we have a Virasoro algebra
together with a $U(1)$ current, then we need two even coordinates
$(z,u)$ in terms of which the evolution operator is given by
$$z^{L_0}u^{J_0}$$
Thus we are naturally led to a 4 dimensional worldvolume theory.  Moreover
in the $N=2$ string
the signature is dictated by the fact that the target has signature $(2,2)$.

In the case at hand we also have a $U(1)$ current algebra
and the geometrization naturally suggests a $(2,2)$ reformulation
of the theory.  This will also `explain' why we have no
oscillatory modes in the compactified (1,1) part of the (10,2) space.
To connect with the usual $(9,1)$ reformulation it is natural
to believe that the wrapping of a (1,1) part of the (2,2) worldvolume
of the F-theory about the compact (1,1) space, leaves us with a
(1,1) string in 10 dimensions, to be identified with type IIB strings.
In such a picture
$(p,q)$ strings \sch\ may arise from
distinct wrappings of the (1,1) part of the
(2,2) worldvolume around the (1,1) compact space.
It seems plausible that the description of this (2,2) worldvolume theory
is related to some of the supersymmetric
self-dual theories proposed in \ref\sieg{W. Siegel,
Phys. Rev. Lett. {\bf 69} (1992) 1493.}\ref\sez{E. Bergshoeff and E.
Sezgin, Phys. Lett. {\bf B292} (1992) 87.}.
Also a prediction of F-theory (and
more precisely the type II string with an extra $U(1)$
on worldsheet)
is the existence of
a peculiar not fully lorentz invariant supergravity theory
in $(10,2)$ dimensions, whose formulation requires the choice
of a null vector--it will be interesting to develop this further.

The idea of a 12 dimensional theory underlying type IIB has
been considered by other physicists:
It was pointed out to us by Sezgin that, quite
independently from $N=2$ string
considerations \ov\ reviewed above,
 it had also been conjectured based on
p-brane/supersymmetry
considerations \ref\du{M. Blencowe and M. Duff, Nucl. Phys.
{\bf B310}(1988) 387}.  We have been told (by Banks and Seiberg) that the
idea of a (10,2) theory underlying the type IIB theory has
also been recently considered \ref\town{P. Townsend et. al., unpublished}.
Also a 12 dimensional Euclidean origin for type IIB was
hinted at recently in \ref\hull{C. Hull, hep-th/9512181}\ though we do not
understand
the relation of those considerations to the ones in the present paper.

\newsec{Compactifications of F-Theory to Lower Dimensions}
The idea of considering lower dimensional
compactifications of F-theory is a
simple generalization of the compactification down to eight dimensions
discussed in section 2.
In reformulating what we found in that case in terms of F-theory,
we should say that we have a compactification of F-theory on
$K3$ manifolds which admit elliptic fibration and where only
the complex moduli of the elliptic fiber is dynamical.
Moreover we have to restrict the moduli so that we
have a fixed $S^2$ sitting in $K3$.
In this reformulation the fact that we got 18 complex moduli
together with one overall Kahler class of $S^2$ is a well known
mathematical fact
for elliptically fibered $K3$ surface.

Now consider a manifold $K$ which admits an elliptic fibration.
We can then consider
compactifications of F-theory on the manifold $K$.
If we compactify further on $K\times S^1$ this theory is on the same
moduli as M-theory compactified on $K$, at least if as in the
discussion of section 2, the adiabatic argument can be trusted.
This relation with the M-theory
will be useful in determining certain general features of the
F-theory
compactification, such as
the number of supersymmetries and the massless spectrum.
In particular
this relation motivates consideration of compactifications
of F-theory on manifolds which preserve a covariantly constant spinor.

Note that upon a further compactification on $S^1$ we get the relation
between type IIA compactification and M-theory. So we conclude
that type IIA compactification on $K$ is on the same moduli
as M-theory on $K\times S^1$ which is on the same moduli
as F-theory on $K\times S^1\times S^1$.  This structure is useful
to keep in mind as we briefly discuss certain compactifications
of F-theory below.

 We will now briefly consider the compactifications
of F-theory on Calabi-Yau threefolds, $G_2$ holonomy and $Spin(7)$ holonomy
manifolds to 6,5 and 4 dimensions respectively.

\subsec{Compactification to 6d}
If we consider compactifications of F-theory on Calabi-Yau
we end up with a theory in 6 dimensions with $N=1$.  To see the number
of supersymmetries note that compactifying further on $T^2$ we should
get $N=2$ in four dimensions, as that would correspond to ordinary
compactifications of type IIA on Calabi-Yau manifolds.  This gives
us a wealth of new models as there are a large number of
Calabi-Yau manifolds which admit elliptic fibrations!  We
are currently analyzing these models \ref\morv{D. Morrison and C. Vafa,
work in progress.}.  Some of these models descend down in four
dimensions to the heterotic/type II dualities found in 4 dimensions
\ref\kv{S. Kachru and C. Vafa, Nucl. Phys. {\bf B450} (1995) 69},
and can be viewed as duals of the corresponding heterotic string
compactifications on $K3$, i.e., a duality in 6-dimensions.  This
is in line with the eight dimensional example discussed
in section 2, where heterotic/type II duality in 6 dimension led
to heterotic/F-theory duality in 8 dimensions.
Other examples will have no analog heterotic duals but may have interesting
type I duals.

Consider an elliptic Calabi-Yau $K$ with hodge numbers $h^{1,1}$ and
$h^{2,1}$.  Let us ask what is the spectrum of the $N=1$ theory in
6 dimensions we will end up with.  In 4 dimensional terms we have
$h^{2,1}+1$ hypermultiplets and $h^{1,1}$ vector multiplets.
Apart from the $N=1$ supergravity multiplet in 6 dimensions
we have tensor multiplets, vector multiplets and hypermultiplets.
Let us denote the number of tensor multiplets by $T$, vector multiplets
by $V$ and hypermultiplets by $H$.  Note that upon dimensional
reduction to $d=4$, vector and tensor multiplets convert to vector
multiplets of $N=2$, and the hypermultiplets remain hypermultiplets.
The $d=6, N=1$ supergravity multiplet leads, in addition to $N=2$
supergravity multiplet,
 to 2 vector multiplets
in $d=4$.  We thus learn that $V+T=h^{1,1}-2$, $H=h^{2,1}+1$.
We can also determine what $T$ and $V$ are separately.
The Calabi-Yau $K$ has $h^{2,1}$ complex deformations.
This fact remains true even if we restrict them to the ones
admitting elliptic fibrations \ref\dmr{D. Morrison, private communication.}.
Together with some other modes
(including $A^+$ constant modes) they make up $h^{2,1}$ of the
hypermultiplets expected.  Let $k$ denote the number of Kahler deformations
of the Calabi-Yau $K$, which does not change the Kahler class
of the elliptic fiber.  Then we get $k$ scalars corresponding to the
Kahler classes.  Note that just as in the eight dimensional example
discussed in section 2 the scalars are not complexified because
the $B$-fields are frozen to be zero.  Out of these $k$ scalars,
the one corresponding to the overall volume of the four manifold
which is the base of the elliptic fibration, together with certain
modes (including the $A^+$ field coming
from the volume of the four manifold) form a hypermultiplet, giving a total
of $h^{2,1}+1$ hypermultiplets accounted for.  The other $k-1$ scalars
are part of $k-1$ tensor multiplets (note that a tensor multiplet
has only one scalar consistent with the freezing of the $B$-fields).
We thus have
$$T=k-1 \qquad V=h^{1,1}-k-1$$
Note that as we vary the moduli and as some of the 7-branes
coincide we end up with an enhanced gauge symmetry
as is by now familiar.

To make the discussion a little more concrete let us consider an example
(which as compactification
to four dimensions was also considered in the context of D-manifolds in \bsvo\
and has been extensively studied in \ref\scho{C. Schoen,
J. fur Math. {\bf 364} (1986) 85.} ).
Consider F-theory
on an elliptic Calabi-Yau where it has two elliptic fibers over $S^2$.
The complex modulus of the first torus is denoted by $\tau_1$ and is identified
with the type IIB complex coupling constant in 10 dimensions.  We denote
the complex modulus of the other torus by $\tau_2$.  Consider a fibration
over $S^2$ where $S^2$ wraps 12 times over each of the two fundamental domains.
We can take the two fibrations determined by
$$y_i^2=x_i^3+f^4_i(z)x_i+f^{6}_i(z)$$
where $i=1,2$ corresponds to the two tori.
The corresponding Calabi-Yau has hodge numbers $h^{1,1}=19,h^{2,1}=19$.
The above fibration describes 19 complex parameters: 24 parameters
from the coefficients of the polynomials, minus 3 from $SL(2)$ and $2$
from independent rescalings which does not affect the tori, leaving
us with $24-3-2=19$.  In this case the number of Kahler parameters,
consistent with preserving {\it one} elliptic curve fibered
is $k=10$.  So from above discussion we have $T=9$ tensor multiplets and $V=8$
vector multiplets\foot{The fact that from geometrical fibrations
of the second torus we get tensor multiplets can also be
seen by the fact that type IIB near an $A_{n-1}$ singularity
is dual to type IIA with $n$-symmetric fivebranes \ogv ,
which gives rise to tensor multiplets on the fivebrane worldvolume \ref\cals{
C.G. Callan, J.A. Harvey and A. Strominger, Nucl. Phys. {\bf B359}
(1991) 611.}.}.  Note that we could have deduced $V=8$ by noting
that the independent complex parameters in the first torus is 12
minus 3 from $SL(2)$ and $1$ from overall rescaling, giving a total
of $8$ independent $7$-brane charges\foot{
Just as we were about to release this paper, a paper
appeared \ref\rsen{A. Sen, hep-th/9602010}\ which describes
an $N=1$ model with the same spectrum as the model discussed
above.  It is quite plausible that the two models are dual to one another.
This is supported further by the fact that both models appear to have
the same enhanced gauge symmetries.}.

\subsec{Compactification to $d=5$}
Compactification of F-theory on an elliptic $G_2$ holonomy
7-manifold leads to a theory in 5 dimensions which has $N=1$ supersymmetry.
These theories will appear in the $d=4,N=1$ moduli of M-theory compactification
on $G_2$ holonomy manifolds to 4 dimensions.  It would be interesting
to study them from this point of view.  Note that the existence
of elliptic $G_2$ and elliptic $Spin(7)$ manifolds has been established
in \ref\joy{D.D. Joyce, {\it Compact Riemannian 7-manifolds
with Holonomy} $G_2$, to appear
in J. Diff. Geom.; {\it Compact Riemannian 8-manifolds with Holonomy}
$Spin(7)$ to appear in Inv. Math.}.  In fact many of the examples
constructed
by Joyce admit elliptic fibration.

\subsec{Compactification to $d=4$}
This is perhaps the most interesting case to consider.  We compactify
F-theory on an elliptic $Spin(7)$ holonomy manifold.  We immediately
run to a puzzle:  By the general arguments above upon a
further $S^1$ compactification to three dimensions this should be
on the same moduli space as M-theory on $Spin(7)$ manifolds, which has
$N=1$ supersymmetry in three dimensions.  However this implies that
the original theory in four dimensions cannot have $N=1$ because
that would have led to $N=2$ in three dimensions.  So the puzzle
seems to be that we are predicting the existence of a theory in four dimensions
which has no supersymmetry but upon compactification on $S^1$ in the limit
of small circle it develops an $N=1$ supersymmetry in three dimensions!
This puzzle is resolved by Witten's observation \witcos\
that $N=1$ supersymmetric
theories in $d=3$ do not exhibit a supersymmetric spectrum and in particular
there would be no obstruction to being connected to a four dimensional
theory with no manifest supersymmetry.  In fact it was proposed in \witcos\
that if a supersymmetric theory in three dimensions is connected to
a non-supersymmetric theory in four dimensions
 this may lead to a resolution of the cosmological constant problem.
What was not clear up to now is whether this beautiful idea
admits a concrete physical realization.
Here we see that not only this idea can be realized but that it is
 {\it generically} the case for
compactifications
of M-theory on {\it elliptic} $Spin(7)$ holonomy manifolds.
Type IIA string compactified to 2 dimensions on an elliptic $Spin(7)$
holonomy manifold is on the same moduli as M-theory compactified
on the same manifold to 3 dimensions which is also on the
same moduli as F-theory compactified on the same manifold to 4 dimensions
(it is quite amusing that at least the uncompactified dimensions
(if not the signature) in each
of these three cases agrees with the dimension of worldvolume
of the corresponding theory).  Note that for compactifications
of type IIA on an eight manifold we have a term generated
\ref\vwone{C. Vafa and E. Witten, Nucl. Phys. {\bf B447}
(1995) 261}\ which should be canceled
by turning on appropriate fields (such as
by turning on string condensates in the type IIA
case or membranes in the M-theory case \ref\vwun{C. Vafa and E. Witten,
unpublished}).  It thus seems that compactifications of strings,
M-theory and F-theory to 2,3 and 4 dimensions on elliptic $Spin(7)$
holonomy manifolds should be studied very intensively
and may hold the key to connecting string theory to the real world.

\vglue 1cm

This work was done while I was visiting Rutgers University
Physics department.  I greatly benefited from the stimulating
research environment there.  In particular I would like to thank
T. Banks and N. Seiberg for participation at the initial
stages of this work and for making many helpful suggestions.
In addition I am grateful to D. Morrison for explaining to me
various mathematical facts about elliptically fibered manifolds.
I would also like to thank M. Bershadsky,
H. Ooguri, V. Sadov and E. Sezgin for valuable discussions.

This research was supported in part by NSF grant PHY-92-18167.

\listrefs
\end